\documentstyle[12pt,aasms4]{article}
\received{1998 Sep 29}
\accepted{1999 Apr 26}
\slugcomment{To appear in: 
Publications of the Astronomical Society of the Pacific}
\lefthead{Nemiroff}
\righthead{Continuous Sky Record}

\begin{document}

\title{Towards a Continuous Record of the Sky}

\author{Robert J. Nemiroff\altaffilmark{1} \& J. Bruce Rafert\altaffilmark{2}}
\affil{Department of Physics, 
Michigan Technological University,
Houghton, MI  49931}

\altaffiltext{1}{nemiroff@mtu.edu} 
\altaffiltext{2}{jbrafert@mtu.edu}

\begin{abstract}
It is currently feasible to start a continuous digital record of the entire sky 
sensitive to any visual magnitude brighter than 15 each night.  Such a 
record could be created with a modest array of small telescopes, which 
collectively generate no more than a few Gigabytes of data daily.  
Alternatively, a few small telescopes could continually re-point to scan and record the entire sky down to any visual magnitude brighter than 15 with a recurrence epoch of at most a few weeks, again always generating less 
than one Gigabyte of data each 
night.  These estimates derive from CCD ability and budgets typical of 
university research projects.  As a prototype, we have developed and are 
utilizing an inexpensive single-telescope system that obtains optical data from 
about 1500 square degrees.  We discuss the general case of creating and 
storing data from a both an epochal survey, where a small number of telescopes 
continually scan the sky, and a continuous survey, composed of a constellation 
of telescopes dedicated each continually inspect a designated section of the 
sky.  We compute specific limitations of canonical surveys in visible light, and 
estimate that all-sky continuous visual light surveys could be sensitive to 
magnitude 20 in a single night by about 2010.  Possible scientific returns of 
continuous and epochal sky surveys include continued monitoring of most known 
variable stars, establishing case histories for variables of future interest, 
uncovering new forms of stellar variability, discovering the brightest cases of 
microlensing, discovering new novae and supernovae, discovering new counterparts 
to gamma-ray bursts, monitoring known Solar System objects, discovering new 
Solar System objects, and discovering objects that might strike the Earth.  
\end{abstract} 

\keywords{stars: variables: general --- instrumentation: detectors --- 
techniques: photometric --- telescopes --- surveys}

\section{ Introduction }

Astronomers have monitored the sky at least as long as history has been 
recorded. Only recently, however, has it become possible to conveniently {\it 
store} such information.  Sky-monitoring ability is developing along fronts in 
energy, angle, brightness, and time.  The energy frontier can be divided into 
the energy band of observation and the energy resolution of observation.  
Similarly, the angular frontier can be divided into the angular band of sky 
observed, and the angular resolution of observation.  Lastly, the temporal 
frontier can be divided into the length of time an object is observed, and the 
temporal resolution of the observation.  The minimum brightness of observation 
is a convolution of many of the above parameters mixed with parameters that 
define the telescope and detector being used, but for convenience will be 
considered a separate quantity here.  Each front carries with it different 
scientific goals and technological obstacles.  

Major strides have been made recently in the creation of static maps of the sky 
over energy bands across the electromagnetic spectrum.  A few notable examples 
include those in the gamma-ray ({\it EGRET}; Fichtel 1996), the X-ray ({\it 
HEAO}; Wood et al. 1984); the ultraviolet ({\it EUVE}; Antia 1993), the optical 
({\it Second Palomar Sky Survey}, Reid et al. 1991; {\it Sloan Digital Sky 
Survey: SDSS)}, Gunn 1995; {\it H-$\alpha$ southern survey}, Gaustad et. al. 
1997), the infrared ({\it Two Micro All Sky Survey: 2MASS}; Skrutskie et al. 
1995), the microwave ({\it COBE}; Bennett et al. 1996), and the radio ({\it 
FIRST}; Becker, White, \& Helfand 1995).  These surveys improved on numerous 
earlier surveys in terms of angular resolution and limiting brightness.

Even in energy bands where most of the sky has been mapped, little of the sky is 
regularly monitored.  To date, large portions of the sky are continuously 
monitored only in the radio and gamma ray bands.  In the radio band, sky 
monitoring is crudely done, but not stored, by anyone who owns a common radio or 
(non-cable) television.  One more sensitive radio sky-monitoring project that is 
being planned is Argus (Dixon 1993).

In the gamma-ray band, sky monitoring began in the 1960s with the launch of the 
Vela satellites.  A changing armada of Solar System satellites has kept 
continuous watch ever since.  One example of a recent, relatively sensitive all-
sky gamma-ray monitor is BATSE onboard the Compton Gamma Ray Observatory.  
BATSE's 8 detectors monitor the sky visible from low Earth orbit in 16 energy 
bands stretching from about 25 keV to about 1 MeV, with angular resolution 
ranging from 1.5 to 30 degrees, and with time resolution ranging from 2 
microseconds to 2.048 seconds (Fishman et al. 1992).  Most of this information 
is stored and publicly available.

In visible light, sky monitoring has been piecemeal. Paczynski (1996) has 
discussed scientific attributes of optical sky monitoring projects doing massive 
photometry.  Small regions of the sky are frequently monitored to study stellar 
and binary star variability.  Stellar and binary star photometry began this way 
last century.  Many of these observations are ``epochal" in the sense that a 
given section of the sky is returned to only after a given epoch of time has 
elapsed.  More recently, fields on the order of degrees have undergone epochal 
monitoring on a daily time scale down to a visual magnitude of about 20 (Udalski 
et al. 1992, Alcock et al. 1993).  These fields have included the LMC and the 
Galactic Bulge through Baade's window, with tens of millions of stars being 
monitored allowing hundreds of candidate microlensing events to be recorded.  

One survey monitoring part of the sky is the All Sky Automated Survey (ASAS; 
Pojmanski 1997, Pojmanski 1998), which monitors 140 square degrees a few times a 
night.  Stardial (McCullough \& Thakkar 1997) scans a strip of sky automatically 
and places images on the internet in near-real time.  The Amateur Sky Survey 
(TASS; Richmond 1997; Richmond et al. 1998) has been scanning the celestial 
equator with CCD cameras since 1996.  Perhaps, though, the most ambitious 
current epochal sky monitor operating at visible wavelengths is the Livermore 
Optical Transient Imaging System (LOTIS; Park et al. 1997; Williams et al. 
1997).  LOTIS currently captures images of roughly 1/4 of the sky, detecting 
stars as dim as 13th magnitude on a daily basis.  An upgraded system dubbed 
Super-LOTIS has recently begun monitoring part of the optical sky every 21 days, 
detecting stars as dim as visual magnitude 19.  Similarly, the Robotic Optical 
Transient Search Experiment (ROTSE, Marshall 1997) has planned capabilities to 
image perhaps 1/4 of the sky to 14th magnitude on a daily basis.

In this paper, the resources necessary to produce epochal and continuous surveys 
of the entire sky in are estimated as a function of limiting brightness.  
Section 2 will discuss the general theoretical concepts that define sky survey 
limits, and show how they apply to a visible light survey.  In \S 3 example 
implementations are discussed for canonical epochal and continuous survey 
characteristics.  In \S 4, we describe a single-telescope prototype system that 
we are using to monitor about 1600 square degrees to visual magnitude 13. In \S 
5, scientific discussion and conclusions will be given.

\section{ Theory of Sky Monitoring }
\subsection{ Constraints }

Realistic continuous sky surveys should have limits defined by the type of 
scientific return desired.  Nevertheless, sky monitoring is subject to practical 
constraints related to the telescope, the detector, data transfer and storage, 
the observing site, and the criteria used for detection. General constraints on 
digital optical sky surveys are discussed, for example, by Kron (1995).

Telescopes are constrained to have a fixed focal length $f$ and a fixed aperture 
radius $r$ (and hence a fixed focal ratio).  Additionally there might be a 
financial limit on the number of identical telescopes $N_{tel}$ that can be 
deployed.

Detector constraints include a given pixel pitch $p$, bit depth $\beta$, pixel 
number $n_{CCD}^2$ (for a square CCD array), and quantum efficiency $e_{CCD}$.  
The CCD and its electronics will give rise to a given dark current noise $D_c$, 
read noise $R_c$, and a finite readout rate $R_{readout}$.

Detection constraints include that at least $\alpha$ pixels exist on the sky for 
each candidate source, and that a given signal to noise ratio $S/N$ is needed to 
provide the desired scientific return.  Time between observations of the same 
piece of sky creates the duration of the epoch of observation $t_{epoch}$.

Storage constraints include a fixed byte size per pixel $b$, a finite amount of 
data storage available and a limited rate at which data can be taken $R_{data}$.  
For example, $R_{data}$ might be limited by the thinnest data pipe, or by the 
maximum amount of write-able disk space available in one night.

Site constraints include fraction $e_{night}$ of allocated time that can be used 
for direct observation of the sky.  Inefficiencies include weather, cloudiness, 
slew time and the time needed to obtain dark frames. 

It will be shown in the subsections below that once a desired limiting 
brightness is determined, the above constraints will combine to define 
appropriate observing system parameters ranging from the telescope focal length 
and the data acquisition rate to the recurrence epochs for epochal and 
continuous sky surveys.

\subsection{ Brightness Limits Survey Pixel Size }

When the number of sources exceeds the number of pixels, the ability to 
associate specific sources with specific flux changes becomes more difficult.  A 
sky survey could proceed even beyond this limit, and even carry significant 
scientific value, but here we will consider this a practical limit.  To insure 
that dim pixels usually surround bright pixels, the number of pixels should be a 
large factor $\alpha$ greater than the number of sources.  An example value for 
$\alpha = 25$.

The number of pixels $N_{pixel}$ that tile the whole sky is a function of pixel 
size.  In general, $N_{pixel} = \Omega_{sky} / \Omega_{pixel}$.  Given that 
$N_{pixel} = \alpha N_{source}$ at the limiting magnitude, then $\alpha 
N_{source} = 4 \pi / \Omega_{pixel}$.  Given square pixels and that pixel 
diameter $\theta_{pixel} = \sqrt{\Omega_{pixel}}$ then
 \begin{equation}
 \theta_{pixel} \sim \sqrt{ \pi \over \alpha N_{source} }.
 \end{equation}
For strongly anisotropic source distributions, $\alpha$ will change with sky 
location. 

At the start of the design phase of a sky-monitoring program at any wavelength, 
the observer's scientific goals might yield a minimum brightness level that is 
desirable.  At this brightness level, the sky surface density of objects at that 
wavelength should be well known, so that the number of pixels on the sky needed 
could be determined

In this paper we will consider a canonical example sky survey of monitoring 
stars in visible light down to a desired limiting magnitude.  For light in the 
visible Johnson $V$ band, Figure 1 shows the expected surface density of stars 
on the sky for a standard Bahcall-Soneira model of our Galaxy (Bahcall \& 
Soneira 1980; Bahcall 1986), as a function of limiting visual magnitude.   Four 
lines depict stellar surface densities at the labeled galactic latitudes.  Note 
that densities near the Galactic plane can exceed those near the Galactic pole 
by more than an order of magnitude.

Figure 1 gives a starting point for the design of a sky monitoring survey.  From 
scientific concerns one chooses a limiting visual magnitude $m_V$.  Given a 
field at Galactic latitude $b$, one can find the sky surface density of stars, 
which can be multiplied by $\alpha$ to give pixel density.  Alternatively, to 
give an all-sky survey uniformity, one might use the average value of $\alpha$ 
for the whole sky.

\subsection{ Pixel Size Limits Telescope Pointings }

If the observer chooses a limiting magnitude $m_V$ for a survey and plans to use 
a CCD with a given number of pixels, $n_{CCD}^2$, then the minimum number of 
separate pointings, $N_{point}$, each of $N_{tel}$ telescopes must make to image 
the entire sky can be determined.  From last section, the number of pixels 
needed over the entire sky was determined solely from $m_V$ to be $N_{pixel}$.  
Now all the pointings of all the telescopes should cover the whole sky so that 
$N_{pixel} \le  N_{tel} N_{point} n_{CCD}^2$.  Solving for the number of 
pointings gives
 \begin{equation}
 N_{point} \ge { N_{pixel} \over
                 n_{CCD}^2 N_{tel} } .
 \end{equation}
A single telescope cannot see more than half the sky from the surface of the 
Earth.  This limits $N_{tel} \ge 2$, and $N_{point} \ge 1$. 

For a canonical survey in visible light, Figure 2 plots the minimum number of 
pointings needed to cover the entire sky, as a function of the survey limiting 
magnitude $m_V$.  Four telescopes are assumed to be operating simultaneously.  
The number of sky pixels is chosen by setting $\alpha = 25$ pixels per star on 
the sky down to $m_V$.  Lines on the plot depict CCD arrays with $n_{CCD}$ of 
1024, 4096, and 16,384.  Throughout this paper we assume $p$ is a constant for 
all $n_{CCD}$.  The flat part of each curve indicates that a single pointing by 
each of the four telescopes would create more than 25 pixels per star. 

Alternatively, for a continuous survey, the number of pointings is fixed at 
$N_{point} = 1$, and the above equation can be solved for the minimum number of 
telescopes needed to view the entire sky simultaneously.  $N_{tel}$ can be 
discerned from Figure 2 by multiplying $N_{point}$ by 4: $4 N_{point}$ then  
corresponds to the minimum number of dedicated telescopes needed to ensure that 
the number of deployed detector pixels ($N_{tel} n_{CCD}^2$) is greater than the 
number of sky pixels ($\alpha N_{star}$).

\subsection{ Pixel Size Limits Data Volume }

The ability to store data could limit the practical extent of a sky monitoring 
project.  Data volume is therefore estimated for epochal and continuous sky 
surveys.  To start, the minimum data obtainable per recurrence epoch is the 
number of sources at the survey limiting brightness multiplied by the number of 
bytes used to record each source: $b N_{source}$.  For our canonical visual 
survey, the dashed line in Figure 3 shows this minimum limit.

The data volume can be written directly in terms of the number of elements in 
CCD array and the number of observations needed.  More specifically, 
\begin{equation}
 {\rm Data} = { b N_{tel} N_{point} n_{CCD}^2 } .
\end{equation} 

In our canonical epochal survey in visible light, $N_{tel} = 4$ and $\alpha = 
25$. For these assumptions, a plot of data volume per sky scan versus $m_V$ is 
given in Figure 3 for our three canonical square CCD arrays.  The flat part of 
each curve indicates magnitudes where a single pointing of all four survey 
telescopes would create the needed number of sky pixels.

Figure 3 indicates that CCD arrays with the smallest number of pixels create the 
least data for sky scans at bright limiting magnitudes.  Sky scans sensitive 
only to stars brighter than magnitude 15 would create only several Gigabytes of 
data, which might be conveniently stored in present day 8-mm tapes.  

At faint magnitudes, the number of pixels in the CCD array is irrelevant.  It 
is, however, much more difficult to store these scans on a single storage medium 
with technology easily available in 1999.  Perhaps when data storage devices 
increase by a factor of about 100, easy data storage of sky scans to magnitude 
20 may be possible.  Were data storage limits to increase by a factor of 1.7 
each year (Moore's Law), it would then be convenient to store such information 
in roughly 9 years, in the year 2008.

\subsection{ Pixel Size Limits Telescope Focal Length, Aperture }

The angular size of a pixel on the sky combines with the physical size of a 
pixel in the detector to define a unique telescope focal length.  Given that 
each pixel of the CCD array has pitch $p$, then the focal length of each 
telescope must be 
 \begin{equation}
 f = { p \over \sqrt{\Omega_{pixel}} } .
 \end{equation}
Since $f$ is directly proportional to $p$, it is straightforward to extrapolate 
this plot to larger values of $p$.  If we further demand that each telescope has 
focal ratio $F = f/(2r)$, we see that the aperture of each telescope must be $r 
= f/(2F)$.  Again, since $r$ is directly proportional to $f$, it is 
straightforward to extrapolate $r$ for different values of $F$.

For our canonical visible-light surveys, a plot of $f$ versus $m_V$ for 
different values of $n_{CCD}$ is given in Figure 4.  As with previous plots, the 
flat part of each curve indicates magnitudes where a single pointing of all four 
epochal survey telescopes would create the needed number of sky pixels. At faint 
magnitudes, as $m_V$ approaches magnitude 20, the focal length of the telescopes 
needed approach one meter.  At bright magnitudes, the focal length can be a 
centimeter or less - much smaller than common with conventional telescopes.  We 
note that camera with such focal lengths are neither impossible nor 
unprecedented in astronomy.  Cameras like this are popularly referred to as 
``wide-angle" or ``fish-eye."  Such a camera would have similarities to the 
human eye.

Every telescope's angular resolution is limited by diffraction.  A convenient 
parameter quantifying an angle where diffraction effects become important is 
$\theta_{diffraction} = 1.22 \lambda / (2 r)$, where $\lambda$ is the average 
wavelength of light being observed.  For an $f/2$ telescope and light with 
wavelength of 5 x 10$^{-7}$ meters, the area in Figure 4 limited by diffraction 
would be below the dashed line.  Figure 4 demonstrates that pixel sizes are so 
large in the optical surveys described that diffraction limits are not generally 
important: stars are assumed to be imaged inside a single pixel with a suitably 
sophisticated optical system design.  Furthermore, blurring by the Earth's 
atmosphere, which would obviate the use of any pixels less than on order 
$\theta_{pixel; min} = 2$ arcseconds, will not be important in any example 
surveys discussed.

\subsection{ Brightness Limits Integration Time }

In a survey in any wavelength band, the objects being surveyed must be viewed 
for a finite time to build up a signal necessary to analyze them.  This amount 
of time is a strong function of the telescope and detector being deployed.  The 
case of stars in visible light is analyzed in some detail below.

The apparent luminosity of a star of magnitude $m$ can be written $I_* = 
I_{\odot} \ 10^{(m_{\odot} - m)/2.5}$ counts cm$^{-2}$ sec$^{-1}$ where the $*$ 
subscript refers to the star and the $\odot$ subscript refers to the Sun. For 
visible light (Johnson $V$), the relation between visual magnitude and
flux detectable over the visual band pass at the top of the Earth's atmosphere can be found from relations in Zombeck (1990) to be $I_{*,space} = 
8.5$ x $10^5$ $10^{-m_V/2.5}$ counts cm$^{-2}$ sec$^{-1}$.  From the Earth's surface, a star visible in a clear sky at zenith will cross one air mass and appear $\Delta m_V = 0.2$ magnitudes more faint (Allen 1976).  Since we are interested in stars at various zenith angles, we will consider that a canonical star crosses two air masses and hence appears 0.4 visual magnitudes more faint:  
\begin{equation}
 I_* = (5.9 \ {\rm x} \ 10^5 \
       {\rm counts \ cm^{-2} \ sec^{-1}} ) \
       10^{-m_V/2.5} .
 \end{equation}
The total counts received from this star are
 \begin{equation}
 C_* = 4 \ r^2 \ t \ e_{CCD} \ I_* = 1.2 \ {\rm x} \ 10^{6} \
      { \left( { r \over 1 \ {\rm cm} } \right)^2 }
      { \left( { t \over 1 \ {\rm sec} } \right) }
      { \left( { e_{CCD} \over 0.5 } \right) }
      { 10^{-m_V/2.5} } ,
 \end{equation}
where $t$ is the duration of the observation, and $e_{CCD}$ is the efficiency of 
the CCD array.  The factor of four is included because it is assumed that the 
CCD array is square, not round.

The brightness of the background sky varies greatly with respect to time and 
location.  Given a sky brightness of $m_{sky}$ in magnitudes arcsec$^{-2}$, the 
apparent luminosity in the Johnson V band pass is given by
 \begin{equation}
 I_{sky} = ( 8.5 \ {\rm x} \ 10^5 \ 
 {\rm counts \ cm^{-2} \ sec^{-1} \ arcsec^{-2} } ) \
 10^{-m_{sky}/2.5} .
 \end{equation}
Note that atmospheric effects that degrade star brightness are 
excluded here.

Now the telescope field of view is  
 \begin{equation}
 \Omega = (4 \pi \ {\rm steradians}) F_{sky} = 
         (5.3 \ {\rm x} \ 10^{11} \ {\rm arcsec}^2) \ F_{sky} 
 \end{equation}
where $F_{sky}$ is the fraction of the sky visible to the telescope.  The field 
of view visible to a single pixel of width $p$ in a telescope of focal length 
$f$ is $\Omega_{pixel} =$ arctan$(p/f) \sim p/f$, assuming the field of view is 
small compared to a radian.  Given a square pixel array of $n_{CCD}^2$ pixels, 
the total field of view of the telescope would be
\begin{equation}
 \Omega =  {n_{CCD}^2 p^2 \over f^2} \sim
           (1.05 \ {\rm steradians}) \ 
           { \left( { n_{CCD} \over 1024} \right)^2 }
           { \left( { p \over 10^{-5} \ {\rm m}} \right)^2 }
           { \left( { 1 \ {\rm cm} \over f} \right)^2 } ,
\end{equation}
and the fraction of sky visible would be $F_{sky} = \Omega / (4 \pi)$.

The total counts received from the sky are then
 \begin{equation}
 C_{sky} = 4 r^2 t \Omega e_{CCD} I_{sky} ,
 \end{equation}
Written in more familiar terms
 \begin{equation}
 C_{sky} = 1.7 \ {\rm x} 10^{6} \
       { \left( { r \over 1 \ {\rm cm} } \right)^2 }
       { \left( { t \over 1 \ {\rm sec} } \right) }
       { \left( { \Omega \over 1 \ {\rm arcsec}^2 } \right) }
       { \left( { e_{CCD} \over 0.5 } \right) }
       { 10^{-m_{sky}/2.5} } .
 \end{equation}

For most observations, the signal is $S = C_*$.  The noise, however, must 
include contributions not only from the background sky counts but also counts 
created during readout and dark time, as well as photon noise.  Dark current in 
the CCD will produce a linearly increasing amount of counts with time: $C_{dark} 
= D t$, where $D$ might be proportional to the ambient temperature.  In sum, the 
noise is given by
 \begin{equation}
 N = \sqrt{ C_* + C_{sky} + C_{read} + C_{dark} } .
 \end{equation}

Scientific observations of a star are usually done above a given signal to noise 
ratio ($S/N$).  Source detection might be accomplished at $S/N = 3$, while 
photometry might demand $S/N = 100$.  The above equations will therefore be 
solved for the time $t$ needed to create a given $S/N$ for a star of magnitude 
$m$. This time will be referred to as $t_m$.  

\subsection{ CCD Saturation Limits Exposure Number }

Detectors cannot integrate for an arbitrary long period, as they will eventually 
saturate.  Detector saturation sensitivity can be characterized by the number of 
bits with which it can record intensity: bit depth $\beta$.  The number of 
corresponding intensity levels is $2^{\beta}$.  As signal is related to noise 
level by the minimum signal to noise ratio that the observer will allow: $S/N$, 
it follows that $S + N = 2^{\beta}$ counts. 

For our canonical survey, we will assume a commonly available detector with a 
depth of 16 bits.  We will also assume that the scientific goals of our 
canonical survey demands a $S/N = 100$.  These attributes combine to limit the 
maximum number of signal counts per single detector integration period to about 
$S \sim 21,400$.  The dynamic range in signals detectable above $S/N = 100$ is 
therefore about 214, which corresponds to a span of about 5.8 magnitudes.  

To reach the limiting magnitude of a survey, multiple exposures might be needed.  
For our canonical survey, we will assume that each exposure lasts until a source 
at the limiting brightness nearly saturates a CCD pixel.  Alternatively, some 
surveys might want to underexpose some frames in an effort to accurately measure 
the luminosity of bright sources.  

After each exposure, the total signal and total background are recalculated, and 
a new $S/N$ is computed for a source at the survey limiting brightness.  After 
$N_{exposure}$ exposures, the desired $S/N$ is reached, and a telescope in an 
epochal survey may be re-pointed.  We note that a telescope in a continuous 
survey will just continue on to new exposures of the same part of the sky, 
building up sensitivity to increasingly dim objects.  In general, the dimmer the 
source, the lower the temporal resolution.  Nevertheless, the continuous survey 
is somewhat optimized for detection of sources brighter than the survey limiting 
magnitude because of source confusion.

The integration time $t_m = t_{integration}$ needed to observe a portion of the 
sky down to the labeled limiting V magnitude $m$ is shown in Figure 5.  Telescope parameters at each limiting magnitude were defined by that magnitude as discussed above for our canonical visual survey.  A detector efficiency of 
$e_{CCD} = 0.5$ was also assumed, and detector read noise $C_{read}$ was taken 
to be negligible.  Note that this integration time is the same for both the 
described epochal and continuous survey.

\subsection{ Efficiencies Limit Survey Recurrence Times }

The timing and duration of sky surveys is limited not only by integration time 
$t_m$, but also by ``inefficiencies" that are inherent in realistic 
observations.   Such inefficiencies include the time needed to take a dark frame 
after each sky exposure $t_{dark}$, the time needed to readout the CCD array 
after each exposure $t_{readout}$, and the time spent waiting for a clear and 
dark sky, here parameterized as $e_{night}$.  Now for a single exposure, 
$t_{readout} = b n_{CCD}^2 / R_{readout}$, where $R_{readout}$ is the readout 
rate of the CCD in bytes sec$^{-1}$.  The total time needed to complete one set 
of observations to the survey limit $m$ would then be 
 \begin{equation}
 t_1 = N_{exposure}(t_m + t_{dark} + 2 t_{readout}) . 
 \end{equation}

For an epochal survey, the total time between observations of the entire sky 
will be $t_{epoch} = e_{night} N_{point} t_1$, where $e_{night}$ is an observing 
efficiency: the fraction of time during which observations can actually occur. 
Combining above equations 
 \begin{equation}
 t_{epoch} = { N_{point} t_1 \over e_{night} } .
 \end{equation}

For our canonical optical survey with the added constraint that $e_{night} = 
0.25$, Figure 6 plots $t_{epoch}$ verses $m_V$.  Note that the ``turnover" point 
where $n_{CCD} =$ 16,384 is favored over $n_{CCD} =$ 4096 occurs at 
approximately magnitude 13 for an epochal survey.

For a continuous sky survey, $t_{continuous}$ is the time it takes to complete 
one set of sky exposures together sensitive to magnitude $m_V$ at a given $S/N$.  
This differs from $t_{epoch}$ since more telescopes are deployed, relaxing the 
demand on any one telescope to a single pointing.  In other words,
 \begin{equation}
 t_{continuous} = { t_1 \over e_{night} } .
 \end{equation}
A plot of $t_{continuous}$ versus $m_V$ is given in Figure 7 for our canonical 
optical survey.

\subsection{ Data Volume and Survey Times Limits Data Rate }

The data volume generated by a single scan of the sky (as depicted in Figure 4 
for our canonical survey) is possibly of less interest than the data 
accumulation {\it rate}.  The data rate is simply the data volume recorded over 
the recurrence time for the survey.  For an epochal survey, 
 \begin{equation}
 R_{epoch} = { {\rm Data} \over t_{epoch} }.  
 \end{equation}
Figure 8 plots a minimum $R_{epoch}$ versus $m_V$ for our canonical survey. 
Inspection of this plot indicates that $R_{epoch}$ actually decreases as $m_V$ 
dims.  This is true so long as data from intermediate exposures is not saved.
 
Inspection of Figure 8 shows that $R_{epoch}$ is at first constant at bright 
magnitudes.  The curve flatness is created by the need for only a single 
exposure and pointing to reach the desired limiting magnitude.  The decrease in 
$R_{epoch}$ at slightly fainter magnitudes is caused by the time needed for 
repeated exposures, although still only a single pointing from each of the four 
telescopes is required.  Finally, near magnitude 20, both multiple pointings to 
different parts of the sky and multiple exposures of each part of the sky are 
required. That $R_{epoch}$ decreases at fainter magnitudes results from the net 
observation time increasing at a greater rate than the data volume.

It appears that the data rate created by any of the considered epochal surveys 
is not enough to challenge even present day data storage media.  The Gigabyte 
storage capacity of modern day 8-mm tapes could reasonably absorb the data 
generated.

The data rate for a continuous sky survey would be 
 \begin{equation}
 R_{continuous} = { {\rm Data} \over t_{continuous} }.  
 \end{equation}
A plot of $R_{continuous}$ versus visual limiting magnitude for our canonical 
visual survey is given in Figure 9.  Note that the data rate is higher than in 
the epochal case, since there are now more than four telescopes accumulating 
data simultaneously.  The combined increase in both data volume and $N_{tel}$ 
will cause $R_{continuous}$ to increase at the very faintest limiting magnitudes 
considered.  At this faint end, the extra telescopes needed using smaller pixel 
arrays would actually take more time to complete their work than fewer 
telescopes mounted with a larger pixel arrays.  Still, for all but the smallest 
pixel array at the faintest limiting magnitude considered, data storage on the 
Gigabyte level per night would be needed, well within the storage capacity of 
modern 8-mm tapes.

\section{ Canonical Implementations }
\subsection{ An Example Epochal Sky Survey }

Attributes of a specific epochal and continuous all-sky survey down to visual 
limiting magnitude of $m_v = 15$ will now be computed as an example of the above 
results.  We will start by assuming telescope and CCD parameters common in 1999.  
Defining attributes will therefore include a CCD array that has $n = 4096^2$ 
pixels each with pitch $p = 10^{-5}$ meters.  The CCD array will be assumed read 
out linearly with $R_{readout} \sim 10^5$ bytes sec$^{-1}$.  We will demand 
$\alpha = 25$ pixels per star for star on the sky brighter than the magnitude 
limit.  Four survey $f/2$ telescopes, operating in parallel, will be assumed.

The magnitude limit of our survey will dictate the minimum pixel density we will 
use on the sky.  For simplicity, we will use a constant pixel density over the 
sky, although it is possible to define surveys where pixel density changes to 
match star density.  Alternatively, it is also possible to use the pixel density 
to determine the limiting survey magnitude, for example integrating longer in 
areas of lower stellar densities.

At magnitude 15, we note that Figure 3 indicates there are about 20 million 
stars on the sky brighter than $m_V = 15$.  Given $\alpha = 25$ pixels per star, 
we might then expect about 500 million pixels to be needed on the sky.  A more 
precise estimate yields $N_{pixel} =$ 5.17 x 10$^8$ pixels will be needed.

Now each of our 4 telescopes has $4096^2$ pixels, yielding 6.7 x 10$^7$ 
pixels in total.  Therefore, at least from the information standpoint, each 
telescope must re-point at least 8 times to record the entire sky.  Assuming 
each pixel generates 4 bytes of data, each sky scan should generate on order 2.1 
Gigabytes.  This information is discernable from Figure 3.

Given the above pixel pitch of $p = 10^{-5}$ m, Equation 4 allows us to compute 
the focal length of the (assumed identical) telescopes.  Each sky pixel would 
take up about $\Omega_{pixel} \sim 4 \pi / N_{pixel} \sim 2.4$ x 10$^{-8}$ 
steradians, so that $f \sim p/\sqrt{\Omega_{pixel}} \sim$ 0.064 meters.  Given 
all telescopes are $f/2$, the diameter of each telescope would also be 0.032 
meters.  This is also evident in Figure 4.

We will (again) assume that the CCD is 16 bits deep, and has detection 
efficiency of $e_{CCD} = 0.5$.  We will assume that the background sky glows at 
21 magnitudes arcsec$^{-2}$.  The exposure time needed to reach visual magnitude 
$m_V = 15$ at a signal to noise $S/N = 100$ is $t_{integration} = 20,598$ seconds (about 5.7 hours), more than the 2974 seconds (about 49.6 minutes) needed to saturate the CCD.  Therefore at least seven separate exposures will be needed.

Integration time is not necessarily the whole story, however.  The time to take 
dark frames, readout times, and the general efficiency of operation would 
combine to determine the true recurrence epoch of this all-sky survey.  Given a 
$R_{readout} = 1.0$ x 10$^5$ bytes per second, a single exposure with a single 
telescope would take $t_{readout} \sim $ 671 seconds to readout.  Assuming that 
a dark frame was required to be taken every time, with the same integration and 
readout times, the total time for one saturated exposure would be about 7290 
seconds, or about 2 hours.  Given 6 near-saturated exposures per pointing and 
one additional exposure calculated to acquire the survey magnitude limit, a complete scan of the sky could be completed in 14 hours.  Given slightly more optimistic limits, a single complete set of sky exposures photometrically sensitive to $m_V=15$ could be achieved in a single night.  Including a general time efficiency (including daytime and cloudiness) of 
$e_{night} \sim 0.25$, however, the total time for a single pointing is $t_1 =$ 50,593 seconds.  Given 8 pointings per telescope needed to cover the entire sky, the recurrence epoch of this survey would run about $t_{epoch} \sim$ 1.6 x 10$^6$ seconds, or under 19 days.  This can be seen in Figure 6.

\subsection{ An Example Continuous Sky Survey }

Let's now assume that we want to create a continuous record of the entire sky. 
Continuous monitoring would accumulate data so long as 
daylight, dark time, and clouds will allow.  (Even daylight records might be of 
interest to meteorologists.)  Each telescope will then be dedicated to a fixed 
area of the sky.  Such dedication might lower systematic errors by reducing 
pointing errors, simplifying data analysis, and reducing telescope complexity.  

An example continuous sky survey sensitive to $m_V = 15$ at $S/N = 100$ would be 
quite similar to the epochal survey discussed above.  A major difference would 
be the number of telescopes deployed.  In the epochal survey, 4 telescopes were 
re-pointed 8 times each.  In the corresponding continuous survey, given the same 
$n_{CCD}$, we would need 32 separate telescopes.  Possibly, the cost of mass 
producing and deploying 32 identical telescopes might be less than 32 times the 
cost of one survey telescope.  Alternatively, a continuous survey might undergo 
incremental implementations, and so not cover the entire sky initially.

The number of pixels generated per sky scan will be the same as with the epochal 
survey.  Given $b=4$ bytes per pixel, the data generated would still be $b 
N_{pixel} \sim 2.1$ Gigabytes.  The focal length of each $f/2$ telescope would 
again be $f =$ 0.064 meters, and the radius $r = $ 0.016 meters.

As above, we will also assume that the CCD is 16 bits deep and has detection 
efficiency of $e_{CCD} = 0.5$, and assume that the sky background is 21 
magnitudes arcsec$^{-2}$.  The main difference between our example epochal and 
continuous surveys comes in the recurrence time.  The extra-dedicated telescopes 
of the continuous survey will cut the time down between re-imaging identical 
patches of the sky.  Since each telescope need do only one pointing instead of 
8, the recurrence time is reduced by a factor of 8.  The recurrence time is 
therefore reduced from about 19 days, to about 2.3 days.  Increases in temporal frequency would require additional telescopes of larger 
aperture, hence, for constant f/ratio, smaller fields of view.

\subsection{ The Potential of Future Sky Monitors } 

What technology development is needed to motivate the creation of a continuous 
record of the entire sky?  This is strongly a function of the wavelength band in 
question.  In general, the development of CCDs and the technological 
replacements for CCDs should be a leading factor.

One potential limiting factor is the readout rate of CCD arrays.  Figure 6 
indicates that at a given limiting magnitude, it is not efficient to use a large 
CCD array, due predominantly to the time it takes to readout the large amount of 
data.  Although physical processes fundamentally limit this time, recent 
progress has been made in reading out CCD sections in parallel.  Were readout 
times reduced significantly or even eliminated, integration time $t_m$ would 
become more closely related to the recurrence time of a sky scan.

Possibly most relevant to the future of astronomical sky surveys appears is the 
growth in size of detector arrays.  Gains in efficiency of CCDs, already 50\% to 
80\%, or decreasing pixel pitch $p$, (which would decrease the photon-capture 
cross-sectional area, requiring longer exposure times to reach the same limiting 
magnitudes), or other chip related technology offer only marginal advances.  
Larger pixels than those commonly available (roughly $30 \mu$m), might be 
useful, however, particularly at brighter magnitudes.

As CCD pixel number increases it becomes possible to record an increasing number 
of stars with a single telescopic system.  Theoretically, an old CCD array could 
just be swapped out for a newer more densely packed array, assuming that the 
optical system were designed with this eventuality in mind.   This would allow a 
survey telescope to go deeper, although both integration time and recurrence 
epoch would necessarily increase.

A simple and perhaps compelling method of predicting the limiting magnitude of 
sky surveys is to assume pixel number as the single greatest limiting feature.  
When pixel number becomes on the order of the number of stars at a given 
magnitude, then a photometric measurement of those stars will be assumed 
feasible in a single night.  Inherent in this assumption is that readout rates 
will stay unimportant as it becomes cheaper to add additional parallel output 
pipes. Lastly, we assume that data storage tapes and drives increase 
sufficiently to allow inexpensive long-term data storage.

McCall \& Corder (1995) have discussed how CCD array technology has increased 
over the past 30 years, even though CCDs did not exist in present form 30 years 
ago.  We here parameterize this as a factor of 1.7 every year.   

In 1999, we will take to be representative of easily obtained arrays $n_{CCD} = 
4096^2 \sim 1.7$ x 10$^7$ elements.  In Figure 3, the dashed line depicts the 
data amount expected were each star to need $b=4$ bytes of data storage.  
Dividing this byte size by $b$, we can see the base number of stars in the 
Bahcall-Soneira model of our Galaxy.  The above pixel element amount therefore 
corresponds to about $10^7$ stars, which occurs at a limiting visual magnitude 
of $m_V \sim 12$.  Given the above pixel factor rate increase, we expect that a 
survey to 15th magnitude in a single night will be possible with common 
resources in between two and three years, and to 20th magnitude in a single 
night will be possible in about 12 years, by about 2011.

\section{ A Prototype System }

To explore both the feasibility of obtaining a continuous record of the entire 
night sky, as well as to familiarize ourselves with realistic operational 
constraints and data handling issues, we have constructed CONtinuous CAMera 
(CONCAM) 1, shown in Figure 10.  CONCAM 1 was built from readily available 
components for a cost of under \$5 K.  It is composed of a 8-mm f/4 fisheye lens, a Meade 8-inch LS-200 mount, and an ST-7 with 765 x 510 pixels each with pitch 6.9 microns x 4.6 microns.  A laptop computer running CCDOPS operates CONCAM 1.  The resulting field of view is 32 degrees x 49 degrees.

CONCAM 1 tracks the sky at the sidereal rate.  Exposures are obtained 
continuously, allowing for readout intervals. Data is stored first on the laptop hard drive but later transferred to read/writable CD-ROMs.  Flat field exposures and dark frames are obtained regularly.  A typical exposure from CONCAM 1 is shown in Figure 11.  Much of the Ursa Major is visible, and the bright star near the frame's center is Dubhe ($\alpha$ UMa).

CONCAM 1 can be described in terms of the quantities defined in Section 2.  The (geometric) mean number of pixels is $n_{CCD} = \sqrt{765 {\rm x} 510} \sim 624$.  Similarly, the mean pixel pitch $p \sim 5.6$ $\mu$m.  The camera sees $\Omega_{tel} = 0.48$ steradians of the sky, giving it a mean angular side length of the field of view of $\theta_{tel} \sim 39$ degrees.  Therefore, at least $N_{tel} = 27$ telescopes would be needed to monitor the complete sky simultaneously.  If only $N_{tel} = 4$ telescopes were deployed, each telescope would need to be repointed a minimum of $N_{point} = 7$ times.  These cameras would divide the entire sky into 1.05 x 10$^7$ pixels.  If we demand $\alpha = 25$ pixels per source, $N_{source} = 4.2$ x 10$^{5}$ sources can be independently monitored, yielding an average sky source density of $\sigma \sim 10.2$ stars deg$^{-2}$.  At this density, source confusion would begin at about $m_V \sim 10$ in directions toward the Galactic plane, at about $m_V \sim 15$ in directions toward the Galactic poles.  

In the near future, CONCAM 2 will be created from an ST-8 CCD camera, and will feature quadruple the static angular coverage of CONCAM 1.  

\section { Discussion and Conclusions }

Technological progress is naturally continuing in a direction that is making a 
continuous record of the entire sky feasible.  Perhaps forethought now will help 
emphasize scientific questions currently at the forefront of current astronomy, 
and help organize observation attributes and data storage methods useful far into the future. 

Continuous sky surveys to bright magnitudes are currently feasible.  Projects 
like LOTIS, ROTSE, and ASAS are already engaging in partial epochal surveys. 
Technology exists today that would enable an all-sky continuous surveys at 
bright visual magnitudes.  For example, there is no fundamental reason why 
astronomers cannot keep running a continuous daily survey of the entire sky 
brighter than magnitude 15, and of millions of stars visible away from the glare 
of the Sun. 

Technology does not usually proceed at a linear (or even logarithmic) pace, as 
technological inventions and procedural innovations often create leaps 
in what is possible.  Alternatively, progress in a field may slow as physical 
limits are approached.  For example, one potentially relevant technological 
innovation includes the development of CMOS detection chips.

Possible scientific returns of a recorded continuous and epochal sky monitoring 
are many and varied, ranging from a better understanding of unusual variable 
stars to discovery of potential doomsday asteroids (Paczynski 1996). It is 
probable that not all-scientific returns can be foreseen.   Nevertheless, other 
potential scientific discoveries of a continuous record of the entire sky might 
include the establishing of case histories for variables of future interest, 
uncovering new forms of stellar variability, discovering the brightest cases of 
microlensing (Nemiroff 1998), discovering new novae and supernovae, discovering 
new counterparts to gamma-ray bursts, monitoring known Solar System objects, and 
discovering new minor-planets.  

Limiting brightness was the starting point for sky survey designs discussed in 
this paper.  Other scientifically chosen observables, such as intrinsic source 
timing or spectral qualities, might also be considered starting points for the 
design of sky surveys.  In many cases the above framework might still prove 
useful.  A computer code from which most of the results in this paper
can be reproduced has been submitted to the Astrophysics Source Code Library (ASCL.NET).

We thank C. Ftaclas and for helpful discussions and comments, and W. Pereira for help with instrumental logistics.  RJN thanks B. Paczynski for 
discussions where he found that he and others already had many of the same ideas 
discussed above.  This research was supported by grants from NASA and the NSF.

\clearpage
\figcaption[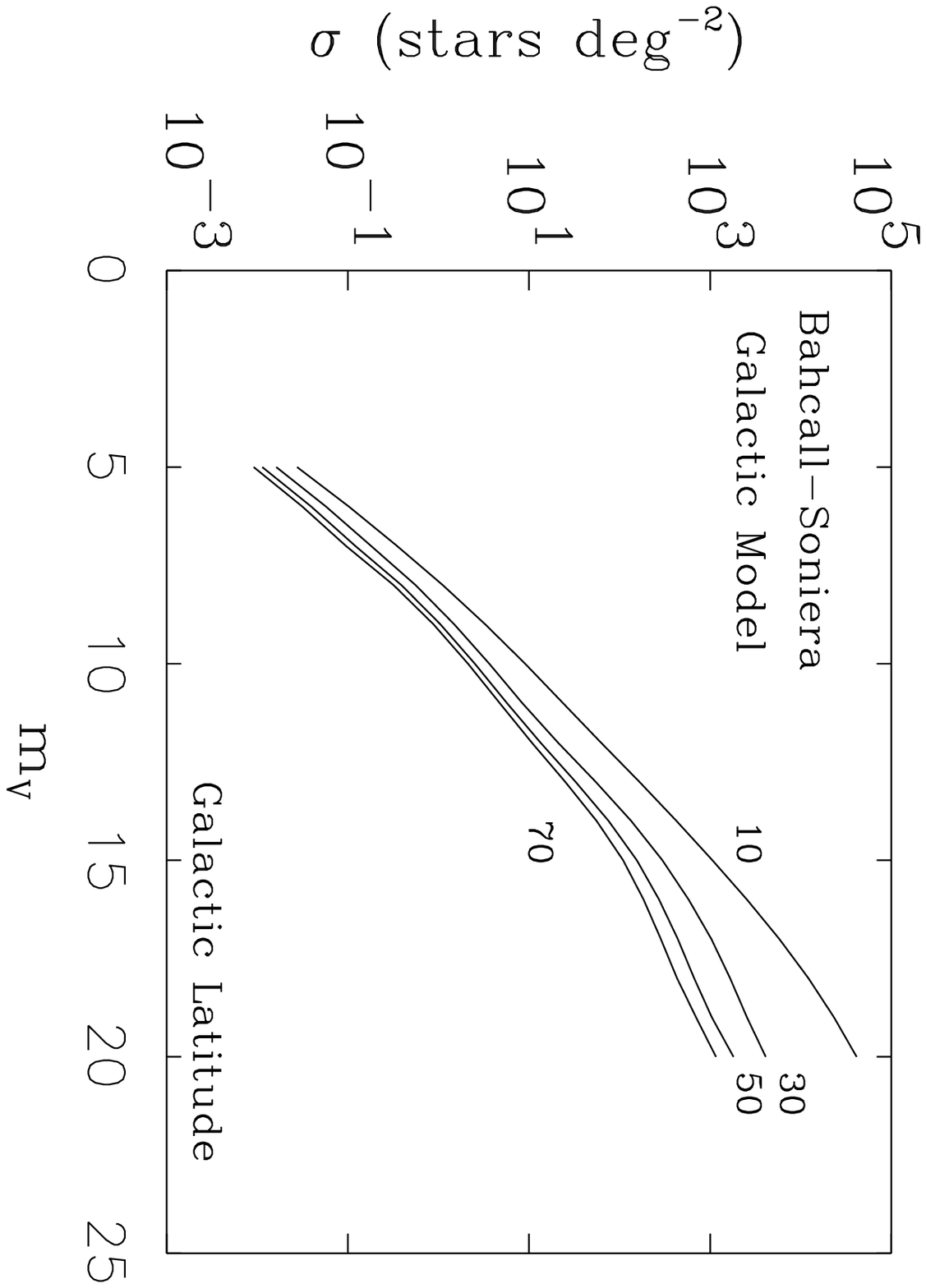]{
The surface density of stars on the sky plotted as a function of limiting visual 
magnitude.  A standard Bahcall-Soneira model for our Galaxy was assumed.  The 
four lines correspond to the labeled Galactic latitude.  When the surface 
density of pixels exceeds the surface density of stars, it becomes difficult to 
monitor stars individually.
\label{fig1}}

\figcaption[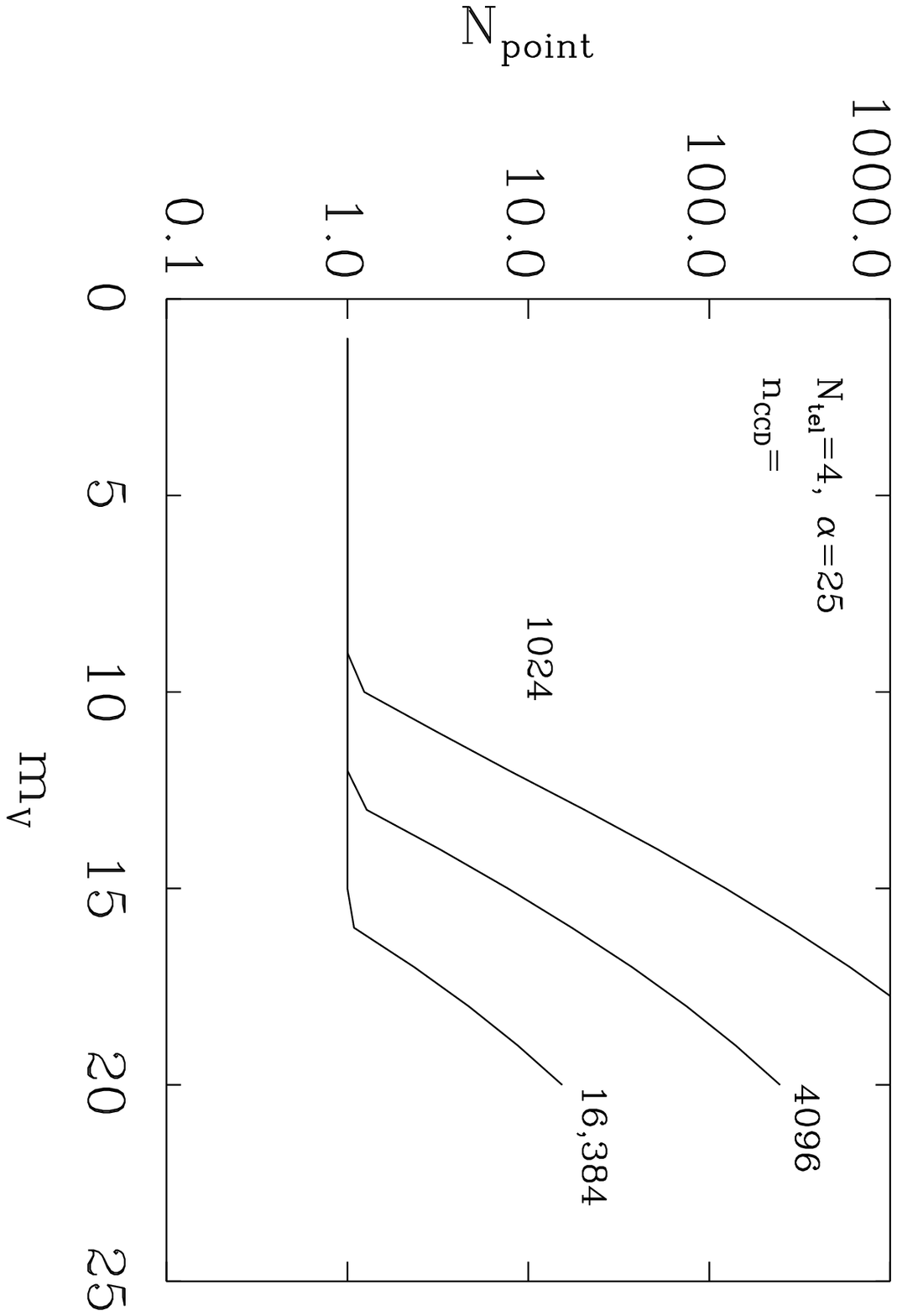]{
The minimum number of telescope pointings needed to image the entire sky as a 
function of the survey limiting magnitude.  Four telescopes and twenty-five 
pixels per disk star are assumed.  The three lines correspond to pointing 
numbers for telescopes utilizing square CCD arrays containing 1024 pixels on a 
side, 4096 pixels, and 16,384.  Larger pixel arrays can tile the sky with the 
needed number of pixels in fewer pointings.  For a continuous sky survey, $4 
N_{point}$ corresponds to the minimum number of dedicated telescopes needed to 
ensure that the number of deployed detector pixels ($N_{tel} n_{CCD}^2$) is 
greater than the number of sky pixels ($\alpha N_{star}$).
\label{fig2}}

\figcaption[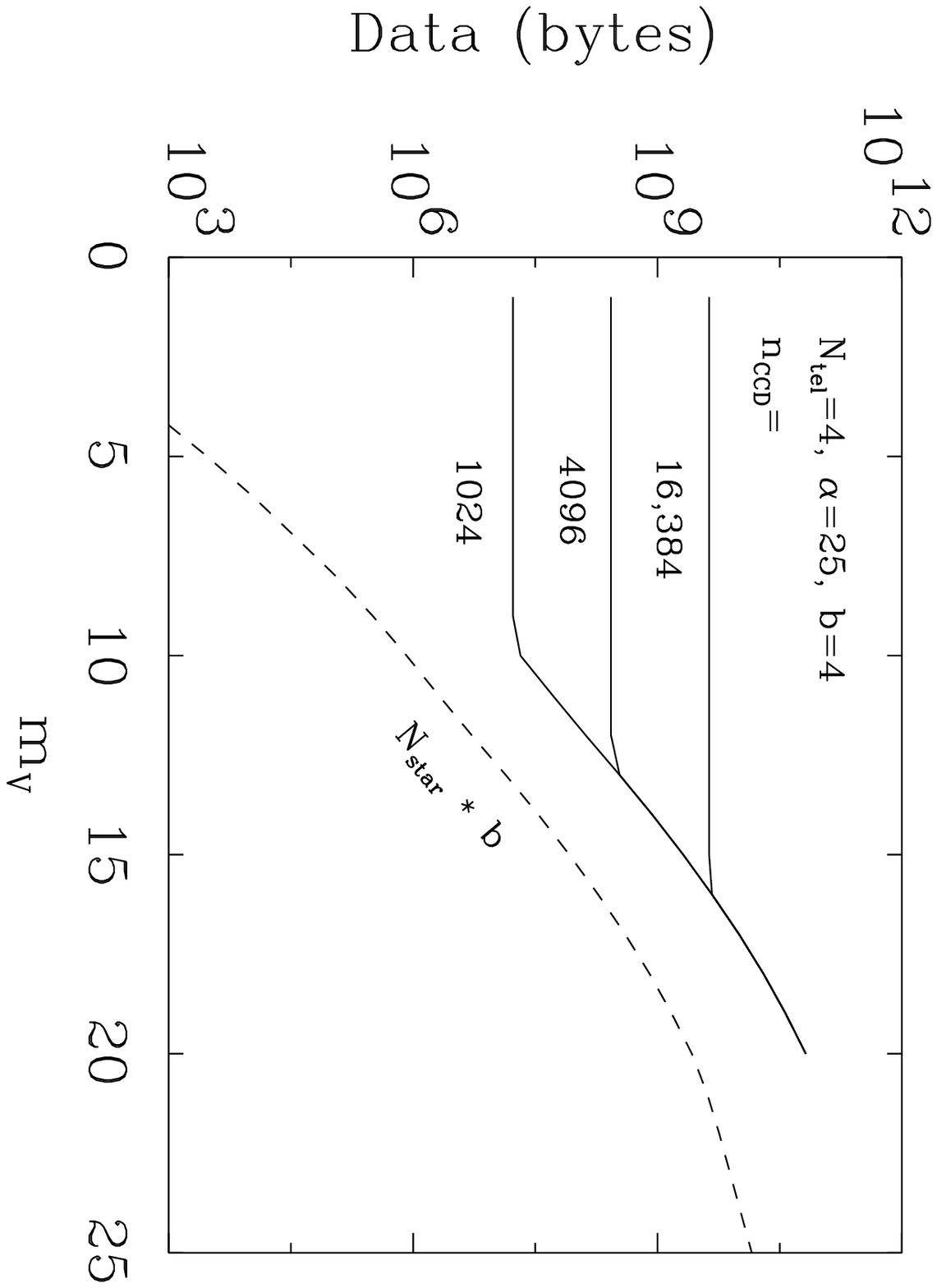]{
The minimum amount of data storage needed to record one image of the entire sky 
as a function of survey limiting magnitude.  The dashed line represents a 
storage minimum, assuming 4 bytes per star.
\label{fig3}}

\figcaption[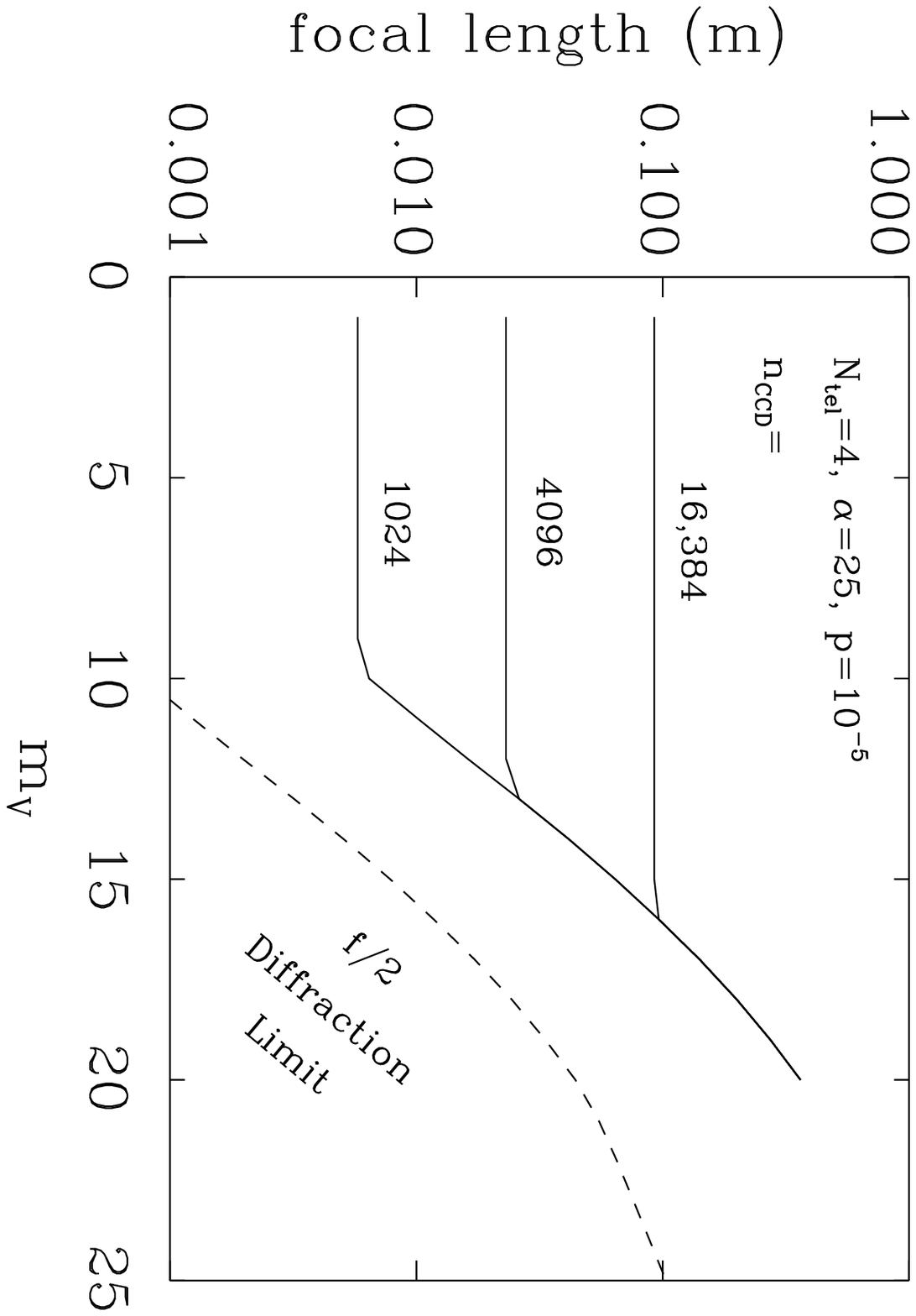]{
For a given angular pixel size on the sky and physical pixel pitch in the 
detector, a unique telescope focal length is implied.  This focal length is 
plotted against the survey limiting magnitude.   The dashed line represents a 
theoretical diffraction limit. 
\label{fig4}}

\figcaption[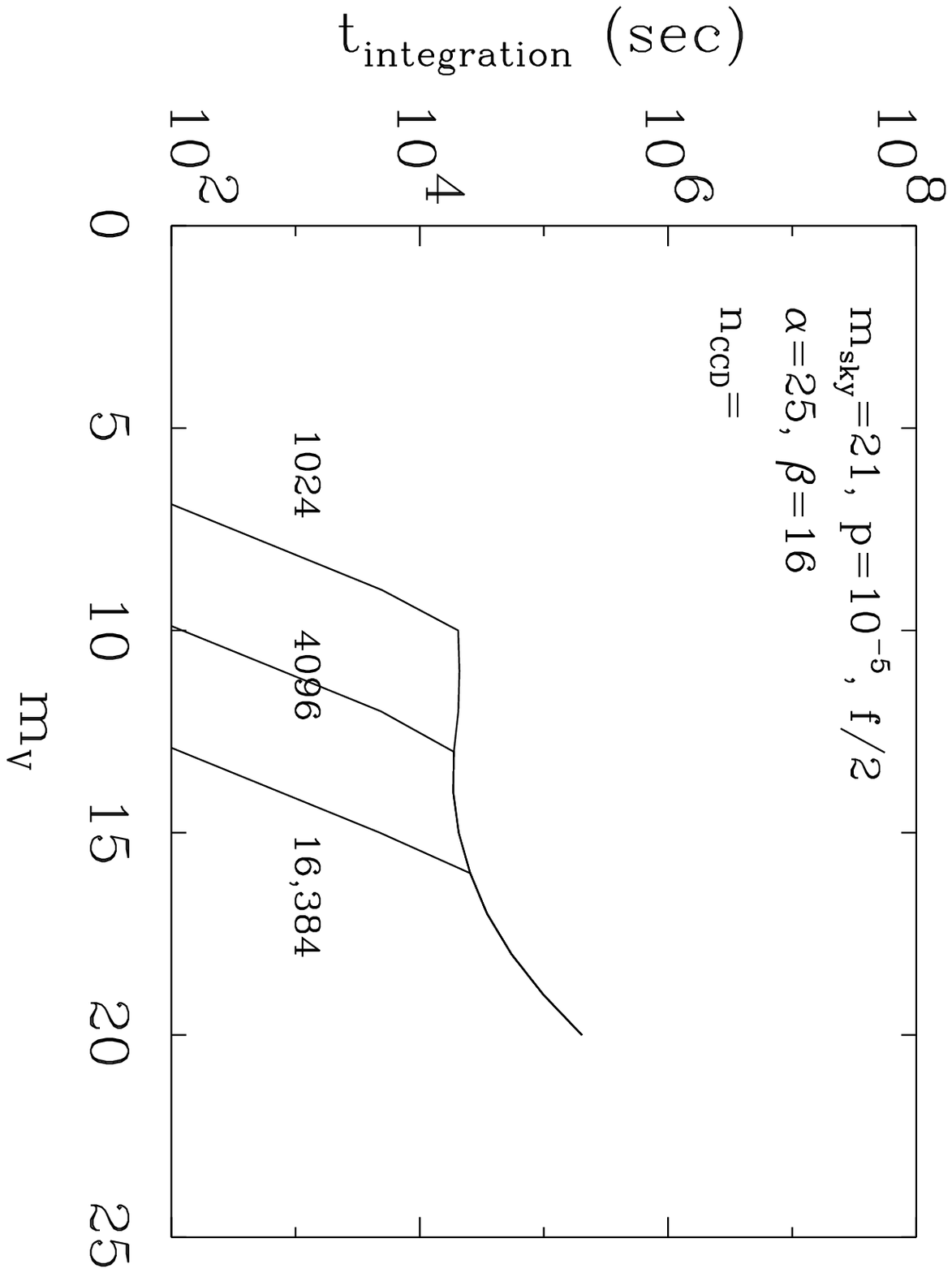]{
The integration time needed to observe a portion of the sky down to the labeled 
limiting magnitude.  An f/2 telescope with focal length depicted in Figure 4 was 
combined with a square pixel arrays of 1024, 4096, and 16,384 elements on each 
side. 
\label{fig5}}

\figcaption[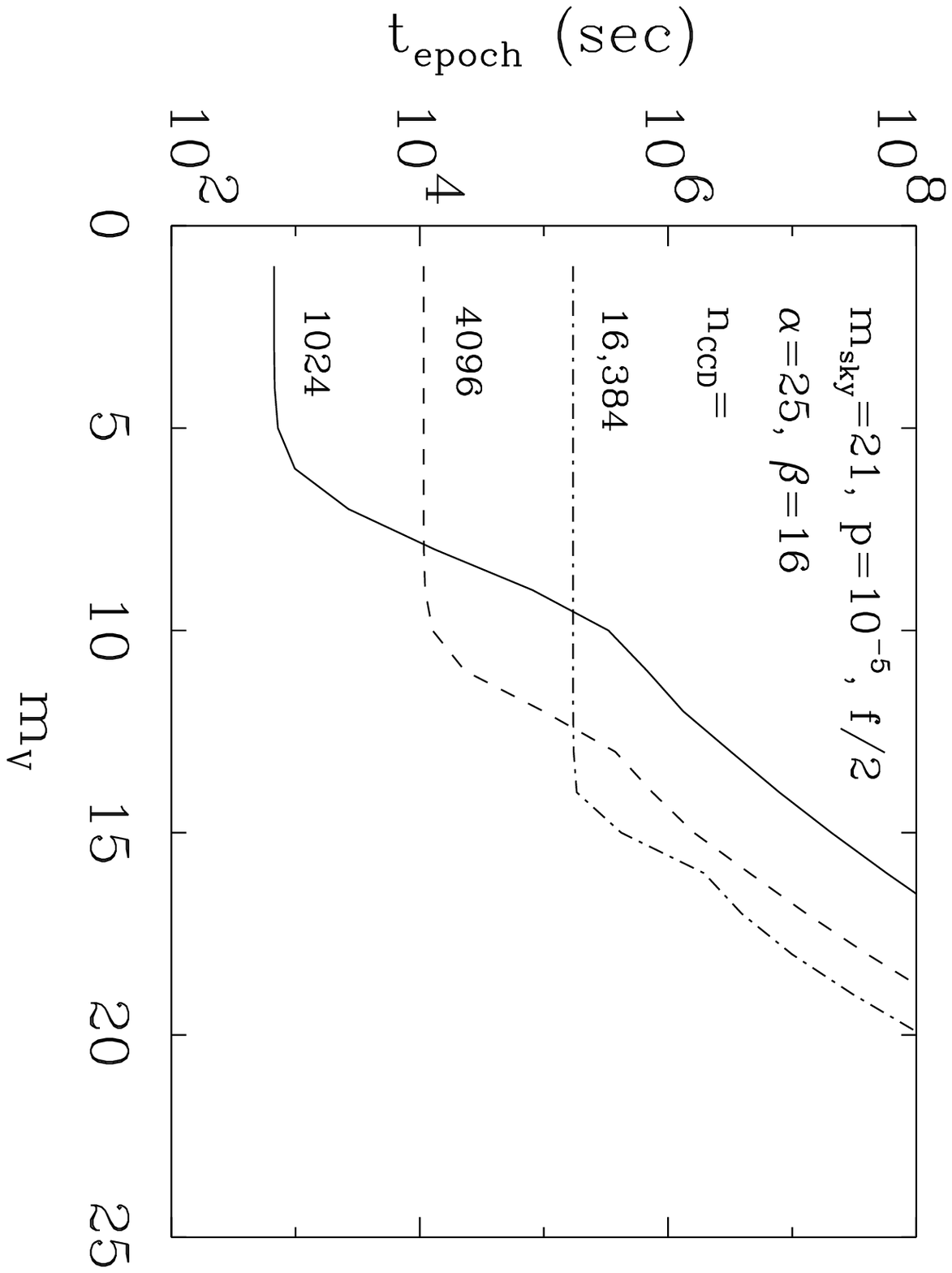]{
The recurrence time of an epochal survey used to monitor the entire sky down to 
the labeled limiting magnitude, using three different CCD arrays.  In the 
epochal survey depicted, 4 telescopes are continually repointed to tile the sky 
to the desired limiting magnitude.  The epoch of recurrence includes data 
readout time, dark frame time, and an estimated efficiency of observation of 25 
\%.  Canonical survey assumptions include 25 pixels per disk star, a background 
sky brightness of 21 magnitudes arcsec$^2$, and a pixel number labeled by each 
curve.  The flat portion of each plot is caused by a single integration being 
able to reach to survey limiting magnitude.
\label{fig6}}

\figcaption[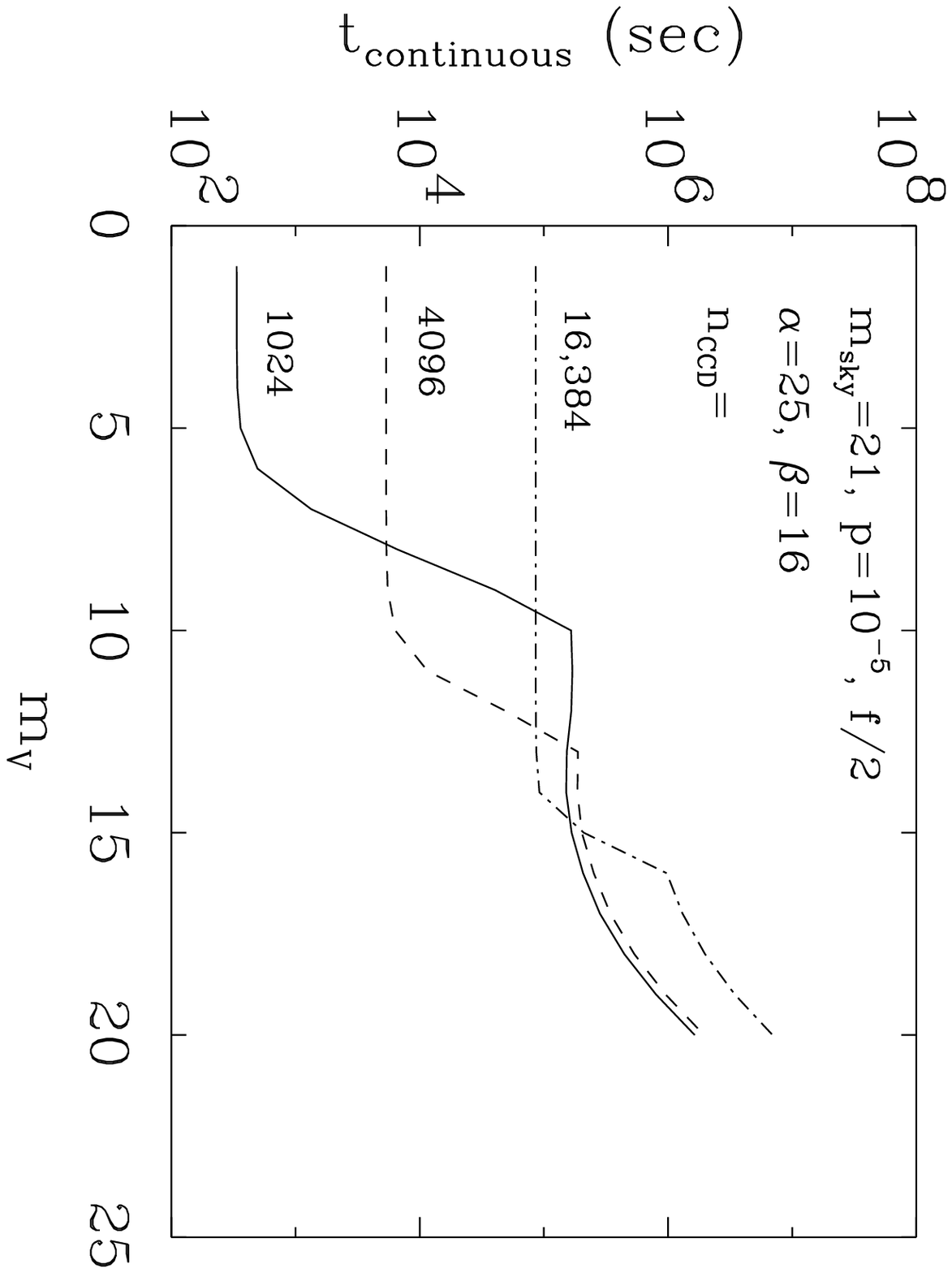]{
The recurrence time of a continuous survey used to monitor the entire sky down 
to the labeled limiting visual magnitude, using three different CCD arrays.  The 
same defining telescope and CCD parameters are assumed as used in Figure 7, with 
the exception that the number of telescopes has been increased so that each 
telescope need not do more than a single pointing.
\label{fig7}}

\figcaption[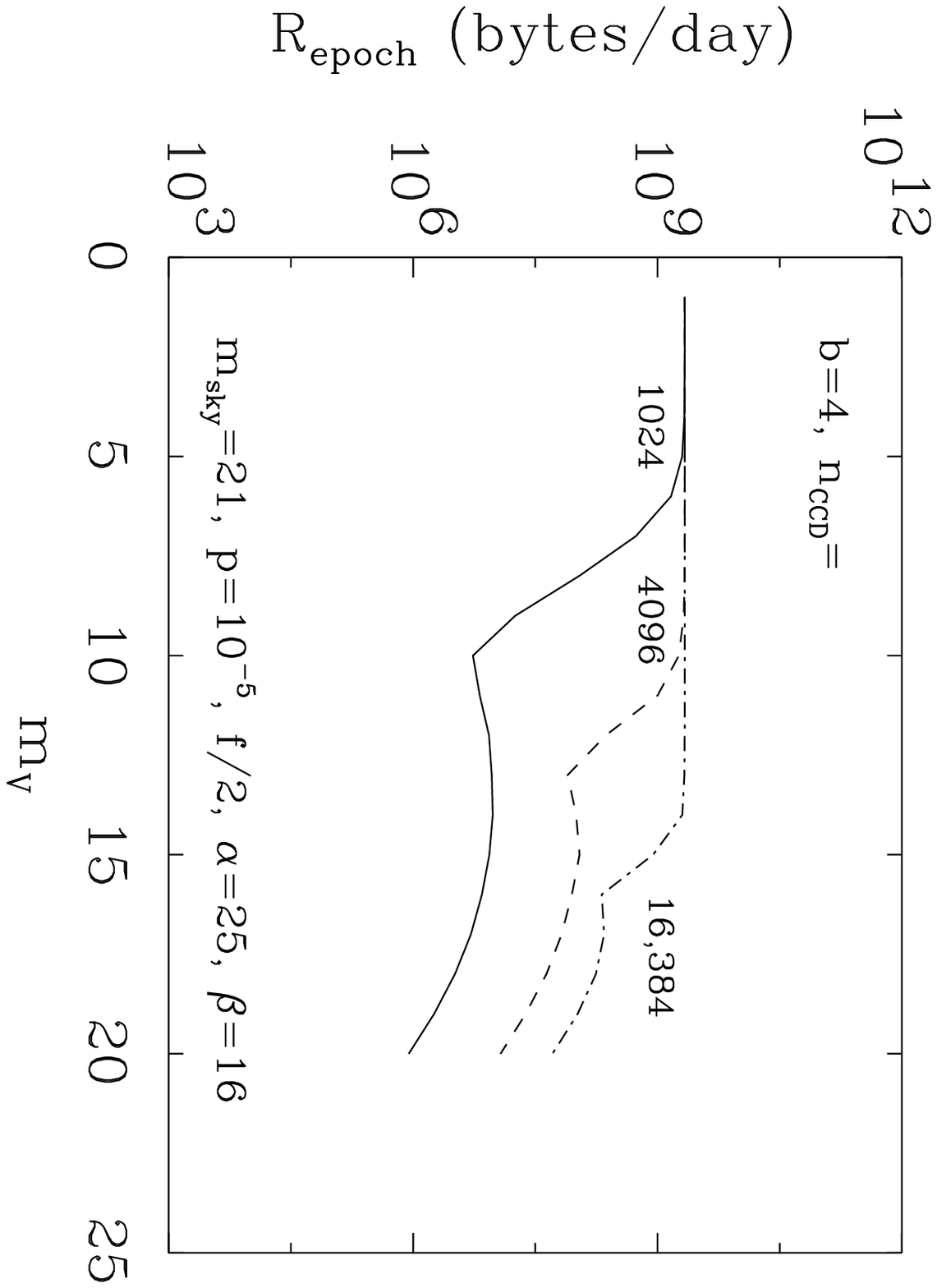]{
The average data rate for an epochal survey of limiting visual magnitude $m_V$.
\label{fig8}}

\figcaption[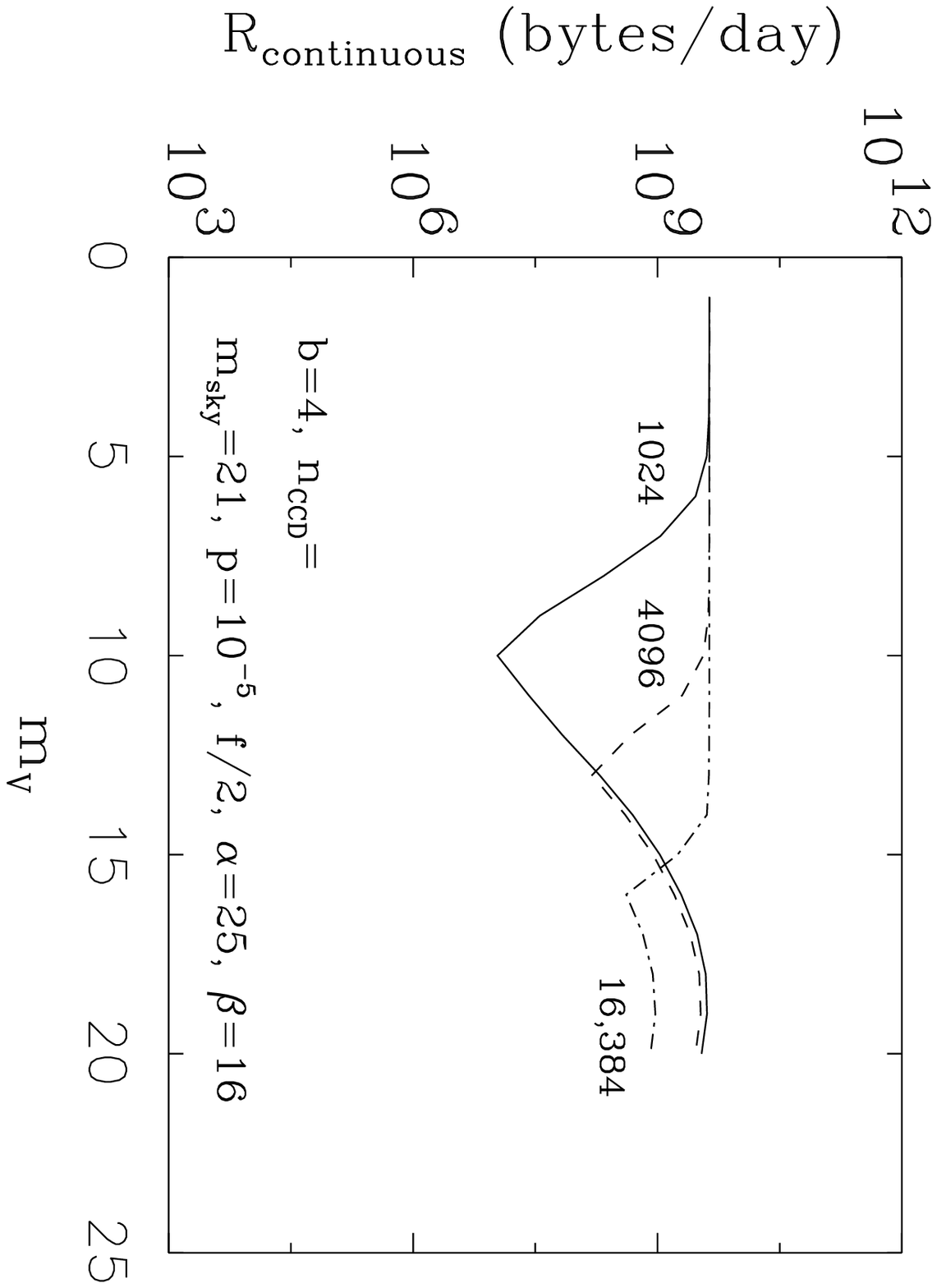]{
The average data rate for a continuous survey of limiting visual magnitude 
$m_V$.
\label{fig9}}

\figcaption[concampic.eps]{
A picture of CONCAM 1, a prototype camera that monitors the sky.
In the LANL preprint version, this figure appears as a seperate gif
file named  concampic.gif.
\label{fig10}}

\figcaption[conskypic.eps]{
A picture of the sky taken by CONCAM 1.  Dubhe, the brightest 
star in Ursa Major, is visible near the field center.
In the LANL preprint version, this figure appears as a seperate gif
file named conskypic.gif.
\label{fig11}}

\clearpage
\plotone{conden.eps}
\clearpage
\plotone{contel.eps}
\clearpage
\plotone{conmem.eps}
\clearpage
\plotone{confoc.eps}
\clearpage
\plotone{contime.eps}
\clearpage
\plotone{conepoch.eps}
\clearpage
\plotone{concont.eps}
\clearpage
\plotone{conerate.eps}
\clearpage
\plotone{concrate.eps}
%
%

\end{document}